\documentclass[useAMS,usenatbib]{mn2e}

\usepackage{amsmath}
\usepackage{graphicx}
\usepackage{color}
\usepackage{ulem}
\usepackage{amssymb}
\usepackage{bm}
\usepackage[T1]{fontenc} 
\usepackage[utf8x]{inputenc}
\usepackage[dvipsnames]{xcolor}

\usepackage{epsf}
\def\msun{{\rm M_{\odot}}}

\def\me{{\dot M_{\rm Edd}}}
\def\le{{L_{\rm Edd}}}

\def\be{{\begin{equation}}}
\def\ee{{\end{equation}}}

\title[Pulsing and Non--Pulsing ULXs] {Pulsing and Non--Pulsing ULXs: the Iceberg Emerges}
\author[Andrew King \& Jean--Pierre Lasota]
{Andrew King$^{1, 2, 3, 4}$ \& Jean--Pierre Lasota$^{4, 5}$\\
$^{1}$ Theoretical Astrophysics Group, School of Physics \& Astronomy, University of Leicester, Leicester LE1 7RH, UK\\
$^{2}$ Astronomical Institute Anton Pannekoek, University of Amsterdam, Science Park 904, 1098 XH Amsterdam, Netherlands\\
$^{3}$ Leiden Observatory, Leiden University, Niels Bohrweg 2, NL-2333 CA Leiden, Netherlands\\
$^{4}$ Institut d'Astrophysique de Paris, CNRS et Sorbonne Universit\'e, UMR 7095, 98bis Bd Arago, 75014 Paris, France\\  
$^{5}$ Nicolaus Copernicus Astronomical Center, Polish Academy of Sciences, ul. Bartycka 18, 00-716 Warsaw, Poland\\     
}

\date{\today}

\volume{000}

\pagerange{\pageref{firstpage}--\pageref{lastpage}} \pubyear{2020}

\begin{document}

\label{firstpage}

\maketitle

\begin{abstract}
We show that ultraluminous X--ray sources (ULXs) with coherent X--ray 
pulsing (PULXs) probably have neutron--star
spin axes significantly misaligned from their 
central accretion discs. Scattering in the funnels collimating their
emission and producing their apparent super--Eddington luminosities 
is the most likely origin of the observed correlation between pulse fraction 
and X--ray photon energy. Pulsing is suppressed in 
systems with the neutron--star spin closely aligned to the inner disc,
explaining why some ULXs show cyclotron features indicating strong 
magnetic fields, but do not pulse. We suggest that alignment 
(or conceivably, field suppression through accretion) generally 
occurs within a fairly 
short fraction of the ULX lifetime, so that most neutron--star ULXs 
become unpulsed. As a result we further suggest that almost all ULXs 
actually have neutron--star accretors, rather than black holes or white 
dwarfs, reflecting their progenitor high--mass X--ray binary and supersoft 
X--ray source populations.
\end{abstract}


\section{Introduction}

Ultraluminous X--ray sources (ULXs) are objects with apparent
(i.e assumed isotropic) luminosities 
$L_{\rm app} \ga 10^{39}\, {\rm erg\, s^{-1}}$, exceeding 
the usual Eddington 
value for stellar--mass black holes, but which do not contain supermassive 
black holes. 

Their discovery around the turn of the century
prompted initial suggestions that they contained black holes with 
`intermediate' masses, i.e above
the then accepted maximum value $\sim 10\msun$ for black holes produced 
by stellar evolution, but below supermassive
\citep[e.g.][]{CM99}. 

But it is now widely accepted \citep[cf][]{Kaaret17}
that most if not all ULXs are stellar--mass binary systems, as first 
suggested by \citet{Kingetal01}. The distinctive features of ULXs
probably result from mass transfer rates which would
produce luminosities significantly above Eddington if they
could be entirely accreted. Systems like this
are the natural outcome of high--mass
X--ray binary (HMXB) evolution, where the accretor is a
neutron star (or in a few cases a stellar--mass black hole) accreting from a 
more massive and largely radiative companion. Once the companion star's
nuclear evolution expands it to the point of filling its Roche lobe, it
transfers mass to the compact star (neutron star or black hole). Because it
puts mass further from the binary mass centre this process shrinks the binary,
increasing the transfer rate. However runaway mass transfer is avoided, as
the largely radiative donor star (of mass $M_2$) 
cannot expand faster than given by its thermal 
timescale $t_{\rm KH}$ \citep[cf][; the basic result goes back to \citet{KW67,KKW67}]{KR99,KB99,KTB2000}
giving transfer rates 
$\dot M \sim M_2/t_{\rm KH}
\sim 10^{-7} - 10^{-5}\msun\,{\rm yr}^{-1}$ or 
even more, well above the Eddington rate
$\me = \le/\eta c^2$ needed to produce $\le$, where 
$\eta \sim 0.1$ the accretion efficiency. Supersoft X--ray sources  
probably have similar
thermal--timescale mass transfer on to white dwarfs, and in some extreme 
cases ($\dot M\ga10^{-5}\msun\,{\rm yr}^{-1}$) may produce
ULXs with white--dwarf accretors 
(King et al., 2001; Fabbiano et al., 2003). Several ULXs
have Be--type companion stars. Here super--Eddington accretion
is not a result of thermal--timescale mass transfer, but instead 
occurs only episodically because the Be--star disc undergoes Kozai--Lidov 
cycles \citep{Martinetal14}. 

Recently,  wind Roche--lobe overflow
has been proposed as a possible mass--transfer mechanism in ULXs 
\citep{Elmellah19,Heidaetal19a,Heidaetal19b}. This automatically
occurs if the companion's
stellar wind has a velocity smaller than the orbital velocities of the binary 
components. But despite the claims in \citet{Elmellah19}, there is no need
to invoke this process to avoid runaway (`unstable') mass transfer once the
stellar photosphere fills the Roche lobe, since this is limited to the thermal--timescale rate discussed above. 

In all cases, the excess transferred
mass is not accreted but ejected, as envisaged by \citet{SS73}.
This explains the survival of the neutron star 
in Cyg X--2 with only modest mass gain after strongly super--Eddington 
accretion (King \& Ritter, 1999).
The excess 
is largely driven away from the 
accretor at a characteristic disc `spherization' radius 
\begin{equation}
R_{\rm sph} \simeq \frac{27}{4}\dot m R_g
\label{eq:sph}
\end{equation}
as a quasispherical wind (of velocity $\sim 0.1c$, King \& Pounds,
2003), which leaves narrow open channels around the accretion disc axis.
(Here $\dot m = \dot M/\me$ is the accretor's Eddington factor,
$\dot M$ is the mass transfer rate from the companion, and
$R_g=GM/c^2$
is the gravitational radius.)
The collimating structure of this wind causes most emitted
radiation to escape along the narrow
channels around the accretion disc axis and makes the true emission 
pattern strongly anisotropic. When viewed along these directions the assumption 
of isotropic emission implies a much larger apparent luminosity
\begin{equation}
L_{\rm app} = L_{\rm sph} \simeq \frac{1}{b} L
\label{eq:lbeam}
\end{equation}
than the true value $L \simeq \le(1+ \ln \dot m)$. Here $b < 1$ is a beaming factor, later found to be approximated by
\begin{equation}
b \simeq \frac{73}{\dot m^2}
\label{eq:b}
\end{equation}
\citep{King09}.
This particular form assumed accretion on to a black hole, and one might
ask if the formula should change in other cases. But
within $R_{\rm sph}$, the accretion rate decreases as 
$\dot M(R) = \dot M (R/R_{\rm sph})$ as gas is progressively expelled. 
This means that the vast bulk of the outflow causing the collimation (`beaming') is 
expelled near $R_{\rm sph}$, 
quite independently of what happens at smaller radii. 
In particular we shall use this form to describe 
accretion in PULXs, where the disc
is disrupted by the neutron--star magnetic field at a radius $R_M < R_{\rm sph}$, and gas inflow is along fieldlines for $R < R_M$. 
We note that our application of this picture to PULXs leads self--consistently to 
values of $R_M$ smaller than $R_{\rm sph}$ and magnetic fields  $\sim 10^{11} - 10^{13}$ G \citep{PULX}.
\begin{table*}
{
\setlength{\tabcolsep}{1pt}
\caption{Derived properties of PULXs (updated Table 2 from \citet{KL19}; see Table \ref{tab:ulx1bis} for observational data.)}
\label{tab:ulx3b}
{\small
\hfill{}
\begin{tabular}{ ||l|||c||c||c||c||c||c||c||c||c||} 
 \hline\hline
 Name &  $\dot m_0$ \ \ \ &$ b$ \ \ &  ${\bm B}\, q^{7/4}m_1^{-1/2}I_{45}^{-3/2}R^3_6$ [G]& ${\bm R_{\rm sph}}m_1^{-1}$ [cm] &  ${\bm R_M} m_1^{-1/3}I_{45}^{-2/3}$ [cm] & ${\bm P_{\rm eq}}q^{-7/6}m_1^{1/3}$ [s] &  ${\bm t_{\rm eq}}$ [yr]$^1$\\
 \hline\hline
 M82 ULX2 &   36  & 0.06 & $9.0\times 10^{10}$ & $3.6\times 10^7$ & $1.0\times 10^7$  &  0.02 & 15600 \\
 \hline
 NGC 7793 P13  &  20 &0.05  &  $2.5\times 10^{11}$ & $2.1\times 10^7$ & $1.6\times 10^7$ &  0.09 & 1386 \\
 \hline
 NGC5907 ULX1 & 91 &0.009 & $2.1\times 10^{13}$ & $9.1\times 10^7$ & $1.1\times 10^8 $  & 1.86 &  0 \\
 \hline
 NGC300 ULX1 & 20 & 0.05  & $1.2\times 10^{12}$$^{\heartsuit}$ & $2.1\times 10^7$ &  $3.2 \times 10^7$ & 0.19& 297 \\
 \hline
 M51 ULX7$^a$ &28 & 0.09 & $1.9\times 10^{11}$  &  $2.9\times 10^7$ & $2.0\times 10^7$ &0.08 & 1337  \\
 \hline
 M51 ULX7$^b$ &28 &0.09  & $6.9\times 10^{9}$  &  $2.9\times 10^7$ & $4.6\times 10^6$ &0.01 & $\sim 10^5$  \\
 \hline
 SMC X-3$^{\rm Be}$   & 18 &0.23 & $2.3\times 10^{10}$ & $1.8 \times 10^7$ &  $7.1 \times 10^6$& 0.006&76621 \\
 \hline
 NGC 2403 ULX$^{\rm Be}$ & 11 & 0.60 & $5.6\times 10^{11}$ & $1.1\times 10^7$ & $2.3 \times 10^7$ & 0.16& 578 \\
 \hline
 Swift J0243.6+6124$^{\rm Be}$  & 14  &0.37 & $1.6\times 10^{11}$  &  $1.4\times 10^7$ & $1.7\times 10^7$   & 0.07 & 2047 \\ 
 \hline
 NGC 1313 ULX$^{\rm Be}$  & 14 & 0.37 & & $ 2.8 \times 10^7$&  & & \\
 \hline
 M51 ULX8& 16 & 0.29 & $\sim 3\times  10^{11}$$^{\clubsuit}$ & $1.6\times 10^7$ &  $2.7 \times 10^7$$^{\spadesuit}$  & non-pulsing \\
\hline\hline.
\end{tabular}}
}
\hfill{}
\vskip 0.2truecm 
 --  Systems with $^{\rm Be}$ superscript have Be--star companions.\\
{$^1$ - calculated using the value of  ${\bm P_{\rm eq}}q^{-7/6}m_1^{1/3}$ from the previous column.}\\
$^{\heartsuit}$--  confirmed by observations \citet{Walton18}.\\
$^{\clubsuit}$-- from observations  \citep{Brightman18,Middleton18}.\\
$^{\spadesuit}$ -- for a $\sim 10^{12}$\,G magnetic field.\\
$^a$ -- for $\dot \nu = 2.8\times 10^{-10}$;
$^b$ -- for $\dot \nu = 3.1\times 10^{-11}$ \citep{Vasilopoulos19}.\\
\end{table*}

In suggesting this picture \citet{Kingetal01} noted its implication that
ULX luminosities alone do not directly tell us the nature of the
accretor, only that its mass must be smaller than the value found by
assuming $L_{\rm sph} = \le$. Accordingly they pointed out that not
all ULXs had to contain black holes, and
predicted that some might instead contain neutron stars or even white dwarfs,
for mass transfer rates
 $\ga 10^{-8}$ or $10^{-5}\msun\,{\rm  yr}^{-1}$ respectively. 
\citep[][identified a possible white dwarf ULX on the basis 
of its unusually soft emission.]{Fabbianoetal03} In the last few years 
observations have revealed at least 10 of the predicted neutron--star 
ULX systems, beginning with ULX M82 X-2 \citep{Bachettietal14},
which has a coherent X--ray periodicity P = 1.37 s, naturally interpreted as
the spin period of an accreting magnetic neutron star 
(see Table \ref{tab:ulx1bis}). Several of these
systems have Be--type companion stars (see also Table \ref{tab:ulx3b}). 

There were initial
suggestions that these pulsing ULXs (PULXs) might somehow evade the 
Eddington limit by having unusually strong (magnetar--like: $B > 10^{13}$\,G)
magnetic fields which reduced 
the electron--scattering cross--section { \citep[see e.g.,][]{Eksi15,Mushtukov15}}, but these 
are contradicted by the $10^{12}$ G fields
found from cyclotron lines seen in NGX300 ULX-1 \citep{Walton18} and 
in the non-pulsing ULX-8 in M51 \citep{Brightman18,Middleton18}.  

\citet{Mushtukov19} also argue that strong outflows observed in PULXs
imply normal pulsar--strength magnetic fields in these objects. These authors
point out that the detection of a strong outflow from a PULX 
is evidence for a relatively weak dipole neutron--star magnetic field.
They also note that an outflow was recently discovered from 
ULX--1 in NGC 300 \citep{Kosec18}, for which \citet{Carpanoetal18} 
deduced a magnetic field of
$B\sim 10^{12}$\,G.  Our model (see below) predicts $B = 1.2 
\times 10^{12}$\,G for this source \citep[see] [and Table \ref{tab:ulx3b}]{KL19}.

\citet[][hereafter KL19]{KL19} noted that the high pulse spinup rates 
$\dot\nu$ observed in the first discovered PULXs
\citep{KL15,KL16,PULX} are a distinctive feature of all pulsing ULXs: 
they are systematically higher than in all other pulsing X--ray binaries (KL19, 
Fig.~1).  Adopting equations (\ref{eq:lbeam}) and (\ref{eq:b}), and
using standard theory for accretion--driven spinup of magnetic 
neutron stars, KL19 found 
the striking result that for all observed PULXs, the magnetospheric 
(Alfv\'en) radius $R_M$, 
where the accretion flow begins to follow fieldlines, is only slightly
smaller than the spherization radius, i.e.
\begin{equation}
R_M \lesssim R_{\rm sph}.
\end{equation}

It  is easy to interpret this relation as a necessary condition for a significant 
pulse fraction to be observable. But it is important to note that it was not 
assumed in advance of the calculation: this necessary condition for 
observable pulsing results simply from applying the theory of \citet{SS73}
for super--Eddington mass supply to the case of a magnetic accretor.

A second significant result of KL19's analysis is
that all the magnetic fields $B$ deduced for
observed PULXs lie in the usual range $10^{11} - 10^{13}$~G for 
all other pulsing X--ray sources (Table 2 in KL19 and Table 
\ref{tab:ulx3b}
in the present paper). There is no need to invoke a new
population of neutron stars with unusually strong magnetic fields to 
explain the PULX population. 

This agrees with the observational 
facts that the measured fields in ULXs are all in the normal X--ray source
range (see above), and that no
magnetar has yet been found to be a member of a binary system. 
In addition there appears to be no easy way of forming
such a system (see, e.g., KL19). For completeness we give here 
in Table \ref{tab:ulx3b} an 
updated version of Table 2 of KL19, adding one new source (M51 ULX7) 
and 
giving the magnetic field values $B$ (rather than the magnetic moments 
$\mu$), together with the deduced beaming factors $b$.

Suggestions that some of the excess accreting matter might be `advected' 
on to the compact accretor's surface \citep[][]{Chashkina19}
are not appropriate for neutron stars -- see e.g. the simulations by
\citet{Takahashi18}) where the advection on to the surface is ``very small"
in a nonmagnetic case,  and by \citet{Kawashima16,Kawashima20}, 
\citet{Inoue20} who show
that emission from the neutron star surface at the bottom of the accretion column 
is negligible.
It is well established that the presence or absence of a critical (sonic) point
above the accretor surface completely changes the nature of the accretion flow 
\citep[][]{Abra10,Lipunova99}.

The results of KL19 offer a tightly self--consistent picture of accretion in 
neutron--star ULXs, which applies to stellar--mass black--hole 
accretors also. A super--Eddington mass supply ($\dot M > \me$)
to a magnetic neutron
star is partially ejected in the way proposed by \citet{SS73}, giving
natural explanations for the distinctive observed features of ULXs and 
PULXs.

Given this agreement, one can attempt a more detailed analysis. PULXs
have significant pulse fractions, up to $\sim 50\%$ \citep{Kaaret17}, and this is sometimes
seen as difficult to make compatible with significant beaming. This paper
discusses this systematically. We will see that the conditions needed for
large pulse fractions strongly constrain the ULX
population.

\section{Pulsing}

X--ray pulse light curve for PULXs are are essentially `sinusoidal', i.e. 
without obvious eclipses, and continuously modulated.  \citet{Mushtukov17,Mushtukov19} 
drew the conclusion that the X--ray emission region must 
in general have an area 
comparable with that of the neutron star, as otherwise there 
would be systems in which this region would either be permanently 
in view, or periodically occulted. The resulting
largely unmodulated emission, or eclipses, 
would conflict with the observed sinusoidal 
character, and  \citet{Mushtukov17,Mushtukov19} 
suggested that the X--rays come
from an optically thick envelope defined by the accretion flow over the
neutron--star magnetosphere. This is physically reasonable, as the 
accretion along fieldlines is a bending hypersonic flow, which must 
therefore shock 
\citep[][KL19]{PULX}. 

But the problem is subtle. The absence of observed eclipses is
much stronger evidence for extended emission regions 
than the lack of unmodulated emission.
Most ULXs do not show pulsing, so it is possible that there are a significant
number of magnetic neutron--star accretors amongst them, changing the
inferred low probability of unmodulated emission. Support for this view
comes from the significant magnetic field inferred from the cyclotron line detected in 
M51 ULX-8 \citep{Brightman18,Middleton18}. We will see that if the X--rays are beamed, both sinusoidal {\it and}
unmodulated emission are quite possible for ULXs where the accretor
is a neutron star with a normal magnetic field.

Mathematically, the pulsed light--curve
problem is identical to that presented by the hard
X--ray light curves of intermediate polars, which are close binaries
where a magnetic white dwarf with a spin period of a few
minutes accretes from
a companion in a close orbit of a few hours. Here also the light curves
are `sinusoidal', and \citet[][hereafter KS84]{KS84}
drew the same conclusion as \citet{Mushtukov17,Mushtukov19}, i.e. that
the hard X--ray emission area occupies a large fraction of the white dwarf
surface. As there is no evidence for unmodulated intermediate polar 
systems, this 
conclusion is not subject to the same possible caveat as mentioned above
for ULX neutron--star 
accretors. To discuss these systems we 
use the analysis of KS84, which classifies 
pulse light curves as functions of the colatitude angle $m$ of the magnetic
field wrt the spin axis of the emitting star 
and the observer inclination $i$ wrt the same axis (for intermediate polars
this is identical with the orbital and disc axes, but this is not true for
neutron--star ULXs -- see the discussion below). 

\begin{figure}
\begin{center}
\includegraphics[width=0.5\textwidth]{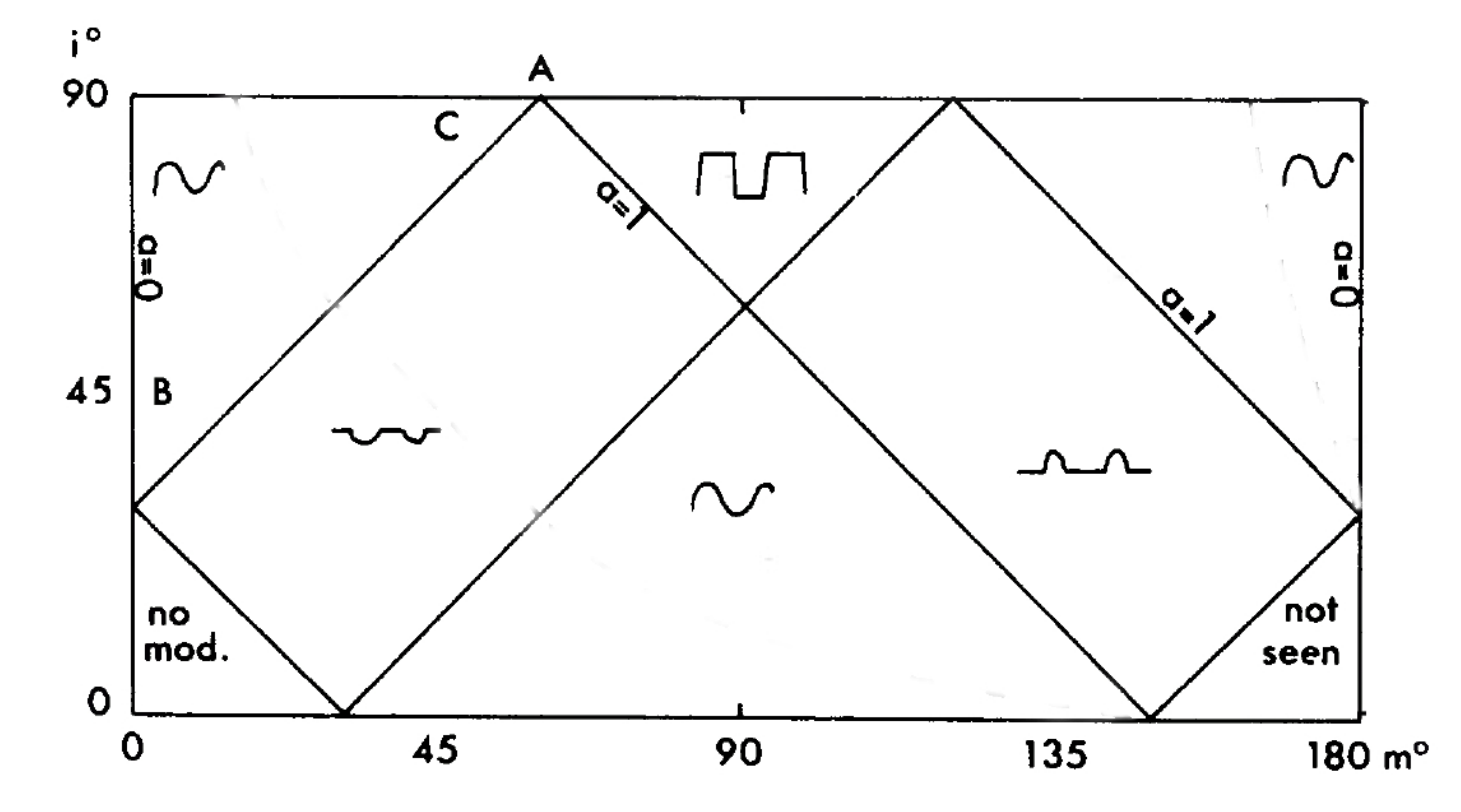}
\caption{ The shapes of X--ray light--curves as functions of angles $i$ (observer 
inclination) and $m$ (magnetic colatitude) for an accreting, spinning
magnetic neutron star with
polecap fraction 0.25, adapted from KS84. The predicted spin 
light--curve shapes are 
indicated in each case by an obvious shorthand. Observed PULX light curves
are continuously modulated (`sinusoidal'), but there may be a large population
of magnetic neutron star ULXs which show {\it no} spin modulation at all. 
 We suggest that observed PULXs are likely to lie in the region 
 $ i \sim 45 - 90^o\ \  {\rm and}\ \  m \sim 0 - 45^o$, where the magnetic axis 
 is strongly misaligned from the spin axis. Alignment of the two axes probably
 occurs in a timescale rather 
 shorter than the ULX lifetime, so that a majority of ULXs
 may contain magnetic neutron stars, but not show pulsing. 
} 
\label{fig:shapes}
\end{center}
\end{figure}

Fig 3 of KS84 shows that the lack of eclipses implies large--area
emission regions. Figure \ref{fig:shapes}, based on Fig. 4 in KS84, shows that sinusoidal pulse 
light--curves can originate from a single `upper' polecap ($m < 90^o$) 
alone if

\noindent
{\it either}
\begin{equation}
 i \sim 45 - 90^o\ \  {\rm and}\ \  m \sim 0 - 45^o 
\label{misal}
\end{equation}

\noindent
{\it or} 

\noindent
\begin{equation}
 i \sim 0 - 45^o \ \ {\rm and}\ \  m \sim 45 - 90^o.
\label{al}
\end{equation}

\begin{figure}
\begin{center}
\includegraphics[width=0.5\textwidth]{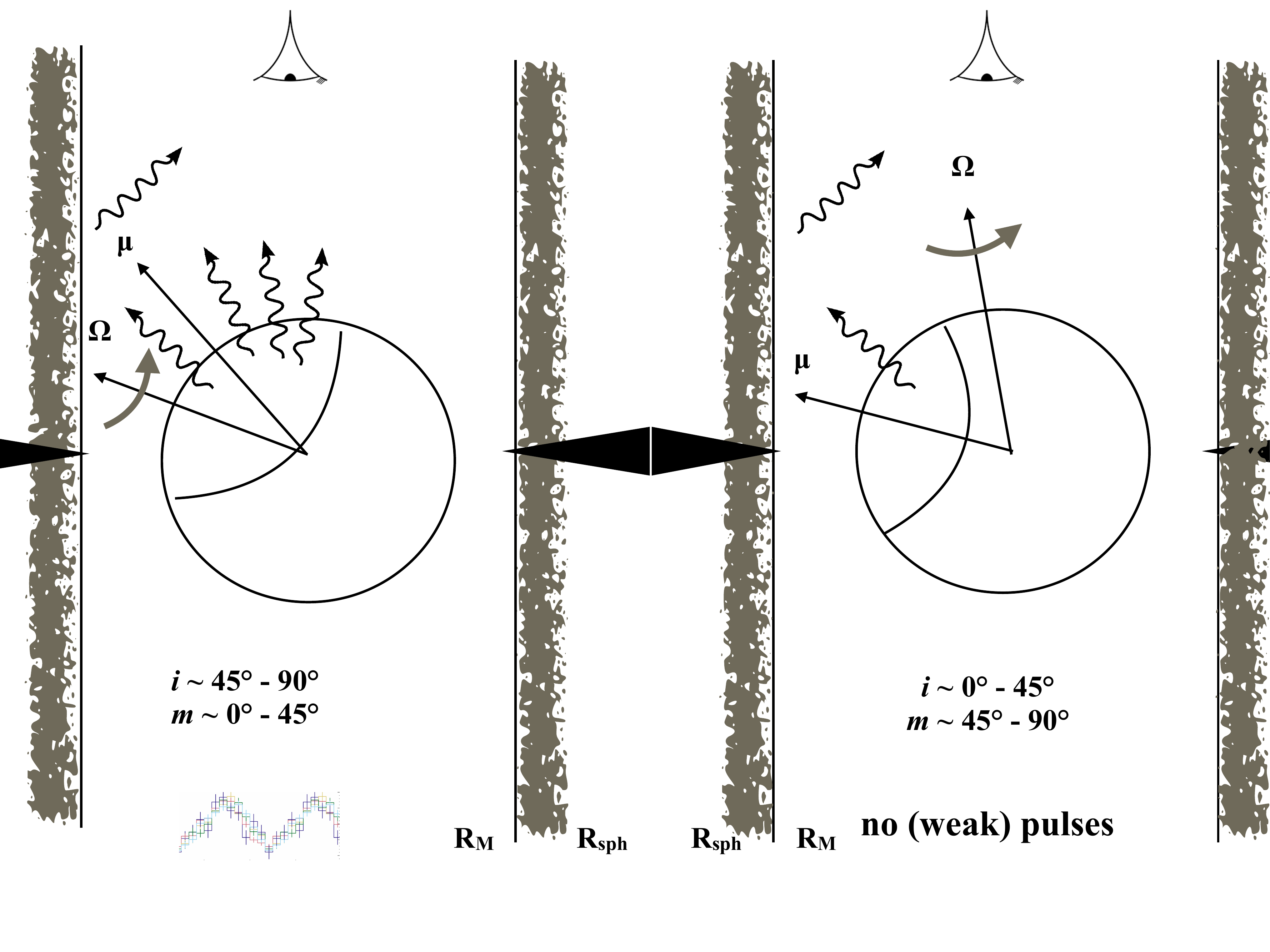}
\caption{Effect of spin orientation on pulsing. Left: the neutron--star spin
and the accretion disc beaming axes are strongly misaligned, so that a 
significant part of the pulsed emission can escape without
scattering. This gives a large pulse fraction. Right: the neutron--star spin 
and central disc axes are substantially
aligned, so that
much of the primary X--ray emission is scattered by the walls
of the beaming `funnel' before escaping. The pulse fraction is
reduced} 
\label{fig:pulses}
\end{center}
\end{figure}

\noindent
(Clearly the incidence of `sinusoidal' light curves is increased further
if instead of a single polecap, a pair of polecaps [$m >90^o$ as 
well as $m < 90^o$] with different
properties are observable at certain spin phases.)
The first case (Eq. \ref{misal}) corresponds 
to a spin axis strongly misaligned from the central disc axis at the spherization radius 
(Fig. \ref{fig:pulses}, left), whereas these two axes are closer to alignment 
in the second case of Eq. \ref{al} (Fig. \ref{fig:pulses}, right) .

Both cases are physically possible. The supernova explosion producing the
neutron star is known from observation \citep[see e.g.,][]{Willems04} to 
be sufficiently 
asymmetric to leave it in general with an initial
spin axis strongly misaligned from 
the  central disc axis. But two processes try to align these axes. The first 
possibility is warping of the central disc into the neutron--star 
spin plane through differential torques (both magnetic and precessional), 
and the second is direct 
accretion of angular momentum (characterized by $R_M$) from an 
unwarped disc lying in the binary orbital plane. The latter seems more 
likely, as any interruption
in the accretion flow means that disc warping has to `start again'. 
Neither process seems likely to be very efficient
in the Be--star PULXs, where accretion is 
barely super--Eddington and confined to short transient episodes.

The outcomes of the two spin orientations  are very different. 

\begin{itemize}

\item If there is strong misalignment of the spin and
beaming axes, a significant part of the pulsed emission can escape without
scattering, giving a large pulse fraction (Fig. \ref{fig:pulses}, left). This must be 
maximal at the
highest X--ray energies, as scattering makes the X--rays both softer and 
less pulsed. This 
correlation of pulse fraction and X--ray energy is well known for observed 
PULXs \citep{Kaaret17}. 

\item If instead there is substantial
alignment of the spin and central disc axes,
much of the primary X--ray emission is scattered by the walls
of the beaming `funnel' before escaping (Fig. \ref{fig:pulses}, right).
Since the light--travel time across
the funnel is usually
comparable with the pulse duration, the pulse amplitude can be
severely reduced. The pulse fraction can of course also be reduced or 
entirely removed if enough matter accretes to weaken the neutron--star 
magnetic field. This presumably happened in the case of Cyg X--2,
which is a survivor of a phase of strongly super--Eddington accretion 
\citep{KR99}. Here the neutron star is not noticeably magnetic,
and has probably gained a few $\times 0.1\msun$ during the 
super--Eddington phase.

\end{itemize}

These outcomes appear to agree well with observations of ULXs. Very
few ULXs show pulsing. But if most ULXs descend either 
from HMXBs once the companion
star fills its Roche lobe, or from Be--star HMXBs, two lines of argument 
suggest that the vast (unpulsed) majority of ULXs must contain neutron stars.
First, almost all HMXBs contain neutron stars rather than black
holes. Second, as remarked by \citet{King09} and \citet{KL16}, 
neutron--star systems are more
super--Eddington and so more beamed than black--hole systems 
with the same mass--transfer rate, and 
so have {\it higher} apparent ULX luminosities.
In addition, there is at least one system  whose 
spectrum shows
a cyclotron feature corresponding to a pulsar--strength magnetic field \citep{Middleton18},
but which does not show pulsing, strongly suggesting that is it an 
aligned accretor. 

The fact that the vast majority of ULXs do not pulse, despite
containing neutron stars, then implies 
that alignment of spin and central disc axes is rapid, and possibly that
field suppression through accretion occurs also. As expected, Be--star
systems are prominent among the PULXs since accretion is relatively weak 
and transient. 

\section{Conclusions}

We have shown that PULXs probably have spin axes significantly 
misaligned from their accretion discs. This geometry naturally reproduces 
the observed
correlation between pulse fraction and X--ray photon energy.
If alignment occurs, pulsing is significantly suppressed, explaining the 
presence of unpulsed ULXs with cyclotron features indicating strong 
magnetic fields. 

We suggest that the timescale for 
spin alignment (or field suppression by accretion) is short compared with
the ULX lifetime, and so that most observed (unpulsed) 
ULXs contain neutron star 
accretors, either with spin and central disc accretion closely aligned,
or with their initial magnetic fields suppressed by accretion.
This is expected if they descend from HMXBs as the 
companion star fills its Roche lobe. Further 
searches for cyclotron features in
unpulsed ULXs offer a possible check of these ideas.

When first recognised as a class, ULXs were widely assumed to 
contain black holes of masses $\ga 100 - 10^4\msun$. The
suggestion here that the great majority actually instead have 
neutron--star accretors, with only a few black holes and white dwarfs,
is a significant shift of the paradigm
\citep[][see also \citet{Mushtukov15}]{KL16, PULX}. We suggest that unpulsed ULXs are the `iceberg' envisaged in the latter paper, and PULXs just the tip of it.

\section*{Acknowledgments}
We thank the referee for pointing out relevant references and helping to 
tighten our arguments.
ARK gratefully acknowledges support from the Institut d'Astrophysique de Paris. JPL
thanks the Nella and Leon Benoziyo Center for Astrophysics of the Weizmann 
Institute
for hospitality and its members, in particular Boaz Katz, for stimulating 
discussions.
This research was supported by the Polish NCN grant No. 2015/19/B/
ST9/01099. 
JPL acknowledges support from the French Space Agency CNES.

\appendix
\section{Observed properties of NSULXs}

In this appendix we present a table with the observed properties of all known neutron-star ULX.
It is an updated version of Table 1 of  \citet{KL19}.
\begin{center}
\begin{table*}
{
\setlength{\tabcolsep}{1pt}
\caption{Observed properties of NSULXs}
\label{tab:ulx1bis}
{\small
\hfill{}
\begin{tabular}{ ||l||c||c||c||c||c||c|} 
 \hline\hline
  Name & $L_X (\rm max)$  [erg\,s$^{-1}$] & $P_s $ [s] &  $\dot \nu$ [s$^{-2}$] & $P_{\rm orb}$ [d]& $M_2$ [$\rm M_{\odot}$] \\
  \hline\hline
  M82 ULX2$^1$ & $2.0 \times 10^{40}$  & 1.37 \ \ & $ 10^{-10}$ \ \ &   2.51 (?) &  $\ga 5.2$ \\
   \hline
  NGC7793 P13$^2$  & $5\times 10^{39}$ & 0.42  \ \  &  $2 \times 10^{-10}$ \ \  &  63.9 & 18--23 (B9I) \\
   \hline
  NGC5907 ULX1$^3$ &  $\sim 10^{41}$ & 1.13 \ \ & $3.8 \times 10^{-9}$ \ \  & 5.3(?) &  \\
   \hline
  NGC300 ULX1$^{4,5}$ & $4.7\times 10^{39}$&  $\sim$31.5 \ \  &  $5.6 \times 10^{-10}$ \ \ &  &  8 - 10 (?)  (RSB) \\
   \hline
   M51 ULX-7$^{6,7}$ & $7\times 10^{39}$ & 2.8 & $2.8 \times 10^{-10}$ ($3.1 \times 10^{-11}$) & $\sim 2$ & $>8  $ \\
  \hline   
  SMC X-3$^{8,9}$&  $ 2.5 \times 10^{39}$& $\sim 7.7$ \ \  & $6.9 \times 10^{-11}$ \ \ & 45.04 & $> 3.7 $(Be ?) \\
   \hline
  NGC 2403 ULX$^{10}$ & $1.2 \times 10^{39} $ &$\sim 18$ \ \ &   $3.4 \times 10^{-10}$ \ \ & 60 -- 100 (?)& (Be ?)\\
   \hline
  Swift J0243.6+6124$^{11}$ &  $ \gtrsim 1.5 \times 10^{39}$ (?) &  9.86  &  $2.2. \times 10^{-10}$ & 28.3 &  (Be ?)  \\
  \hline
  NGC 1313 PULX$^{12}$ & $1.6 \times 10^{39} $ & $\sim 765.6$ \ \ &  & & (Be ?) \\
  \hline
  M51 ULX8$^{13}$  & $2\times 10^{39}$  & \rm NO & \rm NO &  8 -- 400 (?) &  40 (?) \\                                                                                                          
   \hline\hline
\end{tabular}}
}
\hfill{}
\vskip 0.1truecm 
$^1$\citet{Bachettietal14}, $^2$\citet{Furstetal16,Furstetal18,Israeletal17,Motchetal14}; ,\\ $^3$\citet{Israeletal16} 
$^4$\citet{Carpanoetal18}, $^5$\citet{Heidaetal19b},  $^{6}$\citet{RC19}, $^{7}$\citet{Vasilopoulos19}, $^8$\citet{2007ApJ...663..487T}, $^{9}$\citet{Doroshenko18}, $^{10}$\citet{2017A&A...605A..39T}, $^{11}$\citet{Townsend17},$^{12}$\citet{Trudolyubov08}, $^{13}$\citet{Brightman18}.
\vskip 0.1truecm
\end{table*}
\end{center}

\bsp

\label{lastpage}

\end{document}